\documentclass{ceab}

\usepackage{epsfig}
\usepackage{graphicx}
\usepackage{ceabbib}
\usepackage[T1]{fontenc} 

\graphicspath{{./}}
\newcommand{\sect}[1]{Sect.\,\ref{S:#1}}

\newcommand{\fig}[1]{Fig.\,\ref{F:#1}}

\newcommand{\firstfig}[1]{\fig{#1}}

\newcommand{\graphflex}[4][figure]{\begin{#1}\begin{center}#2\end{center}\caption{#4}\label{F:#3}\end{#1}}

\newcommand{\graphwidthflex}[6][figure]{\graphflex[#1]{#5\includegraphics[width=#4]{#2.eps}}{#3}{#6}}
\newcommand{\graphwidth}[4][11.25cm]{\graphwidthflex{#2}{#3}{#1}{}{#4}}
\newcommand{\graphfull}[3]{\graphwidth{#1}{#2}{#3}}

\DeclareRobustCommand*{\unit}[1]{\def~{\,}\ensuremath{\mathrm{\,#1}}}
\usepackage{color}
\definecolor{darkgreen}{rgb}{0,0.45,0}

\begin{document}

\title{Standard 1D solar atmosphere as initial condition for MHD simulations and switch-on effects}
\def\gore{1D solar atmospheric model and switch-on effects}

\author{Philippe-A. \textsc{Bourdin}$^{1,2}$
\vspace{2mm}\\
\it $^1$ Space Research Institute, Austrian Academy of Sciences,\\
Schmiedlstr. 6, A--8042 Graz, Austria\\
\it $^2$ Max-Planck-Institut f{\"u}r Sonnensystemforschung,\\
Justus-von-Liebig-Weg 3, D--37077 G{\"o}ttingen, Germany\\
\tt Bourdin@MPS.mpg.de}

\maketitle

\begin{abstract}
Many applications in Solar physics need a 1D atmospheric model as initial condition or as reference for inversions of observational data.
The VAL atmospheric models are based on observations and are widely used since decades.
Complementary to that, the FAL models implement radiative hydrodynamics and showed the shortcomings of the VAL models since almost equally long time.
In this work, we present a new 1D layered atmosphere that spans not only from the photosphere to the transition region, but from the solar interior up to far in the corona.
We also discuss typical mistakes that are done when switching on simulations based on such an initial condition and show how the initial condition can be equilibrated so that a simulation can start smoothly.
The 1D atmosphere we present here served well as initial condition for HD and MHD simulations and should also be considered as reference data for solving inverse problems.
\end{abstract}

\keywords{Stellar atmospheres - MHD - inversions}

\section{Introduction}

Numerical topics, like the stability of a simulation setup or switch-on effects, are rarely discussed in the astrophysical literature.
While the numerical granularity effects of discrete arithmetics are well studied \citep{Goldberg:1991}, the pseudo-physical consequences of numerical errors are often ignored.
Nonetheless, numerical stability and, equally important, switch-on effects have a strong impact on the physical part of a simulation, be it in space or in time.
Simply said, a simulation is supposed to use only as much computational resources as necessary.
Therefore, the simulation boundaries are usually close to the physical setup.
It is widely known that the spacial boundary conditions have an influence on the physical domain and therefore simulation scientists check for such effects.
What is less known, but equally annoying, is the fact that the same is true for temporal boundaries, in particular switch-on effects.

A typical switch-on effect is for example the ignition of a wave through a sudden release of an external force --- just imagine a string on a guitar that will oscillate long time after striking it.
Instead, if the same string is released smoothly from its deflected position, it can reach the equilibrium position much quicker and will not oscillate afterwards.
The same is true for simulation setups and we will show how oscillations are ignited from a badly chosen model atmosphere and likewise by non-smoothly switching on a simulation.
Avoiding such effects means saving precious computation time and improving the physical results by reduction of methodical artifacts.

We will first present a 1D layered atmosphere suitable to be used from the interior of the Sun up to and including the corona (\sect{solar-atmosphere}).
Then, we discuss the equilibration of a 1D atmospheric model (\sect{equilibration}).
Finally, we demonstrate how a switch-on effect creates spurious oscillations and describe how to prevent such an effect (\sect{switch-on}).

\section{Standard 1D Solar atmosphere\label{S:solar-atmosphere}}

1D atmospheric column models are used to infer basic parameters like density, temperature, magnetic field strength and inclination angle by regularization of inverse problems with spectroscopic observations.
A reference model atmosphere is needed for that and often well-known sample atmospheres are used.
E.g., \cite{Vernazza+al:1981} inferred their model (VAL) from observations, while \cite{Fontenla+al:1993} used radiative HD simulations to get a more realistic profile for the upper chromosphere and transition region (FAL).
Both, VAL and FAL models span only from the photosphere to the transition region.
The steep gradients above the chromosphere are likely to be influenced by the upper boundary condition that is only a few grid points away.
Also the atmospheric stratification is typically fitting only to a quiet Sun area.

A vertical magnetic field is usually also assumed in such 1D atmospheric column models, which is probably only correct for strong magnetic flux concentrations of some \unit{kilo\,Gauss}, like within sunspots and in the center of the main polarities in active regions \citep{Borrero+Kobel:2012}.
And a significant fraction of the field lines become strongly inclined in the upper chromosphere.
The reason for this relative increase of horizontal flux lies in the magnetic field topology, because most of the solar magnetic flux connects back to the Sun already below the corona \citep{Wiegelmann+al:2010}.

Nowadays, we know the solar atmosphere is much more complex and dynamic as was known in the time when the VAL and FAL models were set up.
Therefore it makes sense to use atmospheric models that would not only span higher up into the corona to describe the transition region more realistic, also one should not rely only on a quiet Sun atmosphere for inversions, but take into account a mixed model that would better represent magnetically active areas, as well as inclined field leading to additional variations in the vertical stratification of an atmosphere.

\graphfull{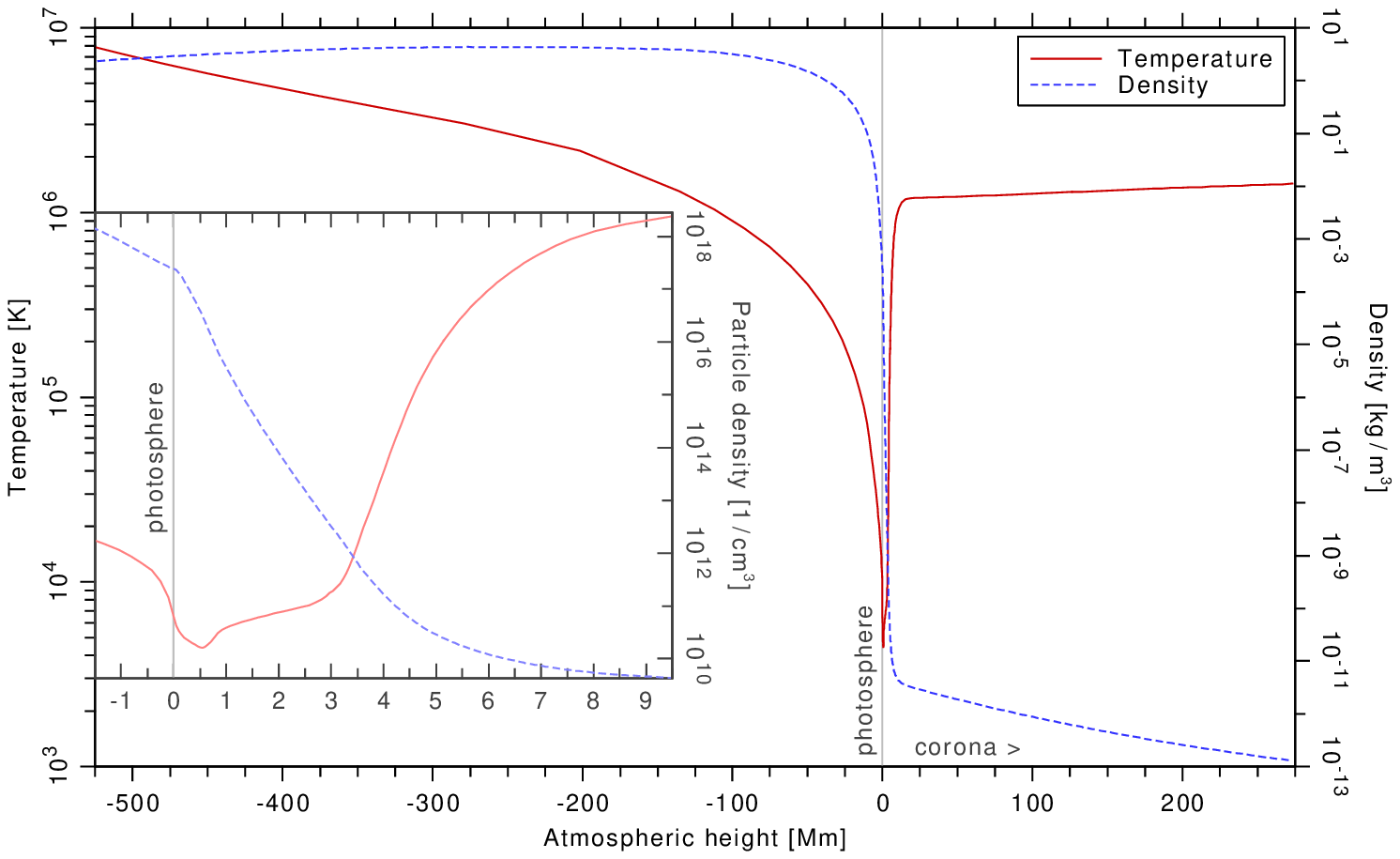}{stratification}{
Combined average stratification of the temperature (red solid line) and the density (blue dashed) in an analytic hydrostatic equilibrium from the solar interior to the corona.
The embedded panel displays a zoom-in from the photosphere up to the lower corona.
The temperature minimum is located here around 500\unit{km} height; the transition region starts around 3.5\unit{Mm}.}

In \firstfig{stratification} we present a combined atmospheric model for the solar interior \citep{Stix:1989} with the FAL data from the photosphere to the upper chromosphere \citep[model C, ][]{Fontenla+al:1993} and a smoothly appended profile for the corona that was inferred from white-light observations \citep{November+al:1996}.
We stretched the middle part taken from \cite{Fontenla+al:1993}, including the chromosphere, so that we can smoothly connect the ends to the profiles of the solar interior and the corona.
This stretching also reflects the averaging over a highly dynamic lower solar atmosphere with a significant fraction of non-vertical magnetic field.
The combined profile still needs to settle into hydrostatic equilibrium because none of the partial profiles are, strictly speaking, in an exact numerical hydrostatic equilibrium on the discrete simulation grid, as we discuss in the next section.

\section{Equilibration of an initial condition\label{S:equilibration}}

Any atmospheric model needs to find an equilibrium state before it can be used as initial condition.
Even though one could in principle start with either a temperature or density profile and analytically calculate the hydrostatic counterpart, this would not mean that the generated atmospheric profile is also in an exact hydrostatic equilibrium within a new simulation setup.
The reason is that a discrete realization of any profile on the grid of a new simulation setup is most probably slightly different than the original profile, even if the original data would be available in its raw form, which it is usually not.
Be it rounding errors, inexact discretization, or any sort of interpolation --- there will be a slight deviation from equilibrium causing a restoration force that will suddenly be released when the simulation starts.
But, more importantly, even if one has at hand the exact analytic solution for an atmosphere in equilibrium, the numerical derivatives are inexact and therefore inconsistent with the analytical ones.
This means that the model again sees a deviation from the exact numerical equilibrium that will have to settle by relaxation motions in some way, usually igniting waves or even shock waves.
Such hydrodynamic disturbances might only dissipate on very long time scales.
And we want to avoid, by all means, such switch-on effects, as well as that pseudo-physical disturbances stay present during the physical part of a simulation.

In \firstfig{hydrostatic_switch-on} we show the time evolution of the vertical velocities of a 1D stratified hydrodynamic atmosphere.
As initial condition we use here the hydrostatic composite of the atmospheric temperature and density profile from \sect{solar-atmosphere} representing the solar atmosphere from the photosphere up to well in the corona.
The top boundary condition is implemented here as a closed boundary, so that waves are to a large part reflected back.
We do this here on purpose to demonstrate the persistent propagation of such shock waves through the simulation for long time that can in some cases even be longer than the main physical part of the simulation is intended to last.

\graphfull{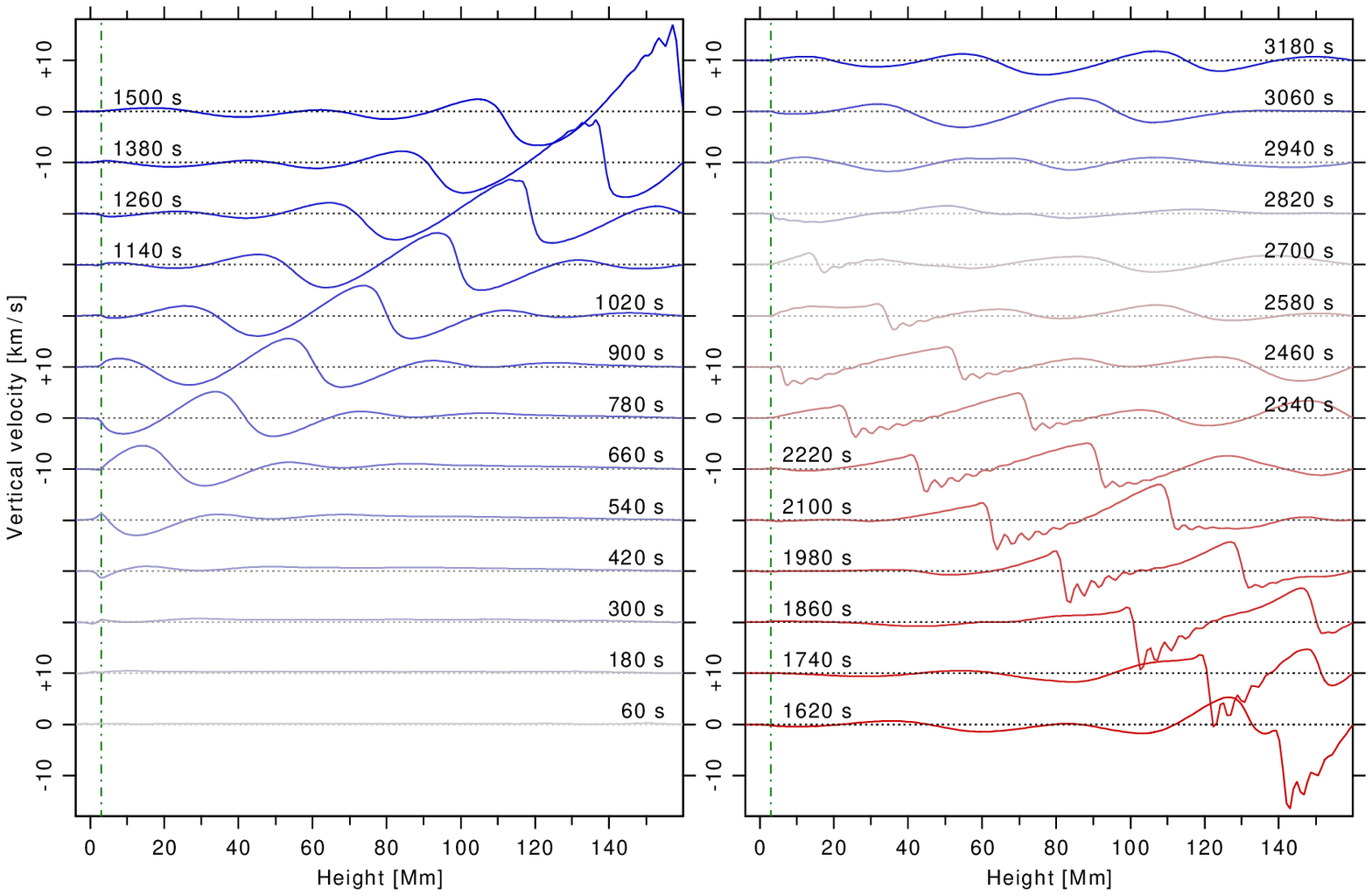}{hydrostatic_switch-on}{
Time evolution of the vertical velocities in a 1D hydrodynamic model atmosphere.
The left and right panels display the two halves (0--1500\unit{s} and 1620--3180\unit{s}) of the simulation run.
For clarity, we shifted each snapshot by 10\unit{km/s} in the vertical direction.
The horizontal gray dotted line indicates the zero position of each velocity profile.
The color indicates the main propagation direction of the shock waves created by the initial condition, where blue is upwards and red is downwards propagation.
The vertical green dash-dotted line is located at the position of the upper end of the chromosphere that is here at 3\unit{Mm} height.}

We find in \fig{hydrostatic_switch-on} that after about 300\unit{s} the slight deviations from the photosphere have propagated to the upper end of the chromosphere, where the density is becoming much lower than in the photosphere (see \fig{stratification}).
This density decrease leads to an increasing amplitude in the vertical velocity of the relaxation motion to account for mass and momentum conservation of the propagating wave.
In the snapshots from 300--1020\unit{s} we see multiple oscillations at transition region height.
These drive then strong waves that are gaining in amplitude when the density drops further with increasing height in the corona.
We also see how a clear shock-wave front forms in the snapshots 1140--1500\unit{s}.

After reaching the top (left panel in \fig{hydrostatic_switch-on}), the relaxation wave is reflected back downwards, flips over due to the asymmetric boundary condition, and interacts also with the still upwards propagating following oscillations (1620--1740\unit{s}).
Later, in the snapshots 1860--2340\unit{s}, the shock-wave amplitude decreases again due to the increase in density that the wave sees when propagating towards the solar surface.
Furthermore, the slight velocity damping that we apply for numerical stability, as well as the viscosity, are both slowly dissipating the wave.

When hitting the steep density gradient in the transition region, the wave is again reflected back upwards at 2460\unit{s}.
The reflected upward wave now lost the shock front character due to the dispersive behavior around the transition region, which acts in a way as a "soft" boundary condition.
After 2820\unit{s}, also the second shock wave is reflected and we see how the disturbances travel again towards the upper boundary.

Comparing with the simulation of \cite{Bourdin+al:2013_overview}, that was the first to show a fundamental match to observations of a coronal loops system above an active region, the analyzed snapshot was obtained after about 3780\unit{s} with no additional damping after the first 2400\unit{s}.
Needless to say, that such strong relaxation motions in the vertical velocities of some 5\unit{km/s} would have completely destroyed the obtained results in that study, where the plasma flow motions within the coronal loops had to be accurate to about some 1--2\unit{m/s} to reproduce observed Doppler red- and blueshifts.

To avoid such relaxation motions from non-equilibrated initial conditions, one should use an atmospheric profile in a supposed hydrostatic equilibrium and fit it in a 1D hydrodynamic atmospheric column to settle these relaxation motions with low computational demand until they have reached an amplitude that would be acceptable as error in the productive 3D simulation run.
To achieve that, one need to use the same grid, the same diffusion parameters, and the same boundary conditions as one likes to have in the production run.

Also one should keep in mind that a hydrodynamic equilibrium also includes the vertical velocities.
Resetting the velocities to zero in a (mostly) relaxed atmosphere, even though they might be tiny, will again disturb the obtained hydrodynamic equilibrium and will again lead to instant relaxation motion and subsequent shock waves propagating through the model, when starting the production run.
Over-damping the relaxation motions is not an option, because once the damping is released in the production run, the same effect will appear.
Therefore, one should transfer not only the temperature and density profiles to the production run, but also the matching vertical velocity profile.

\section{Switch-on effects\label{S:switch-on}}

After having settled the initial condition, one often wants to driver a simulation with some external forcing, e.g. by mixing motions on one boundary.
In \firstfig{driver_switch-on} we show a 3D MHD setup with a turbulent flow.
The mixing we apply through driving velocities at the left boundary that remain within the plane of that boundary.
A flow is generated that propagates to the right and leaves the simulation domain on the right boundary that we open for outflows.

\graphfull{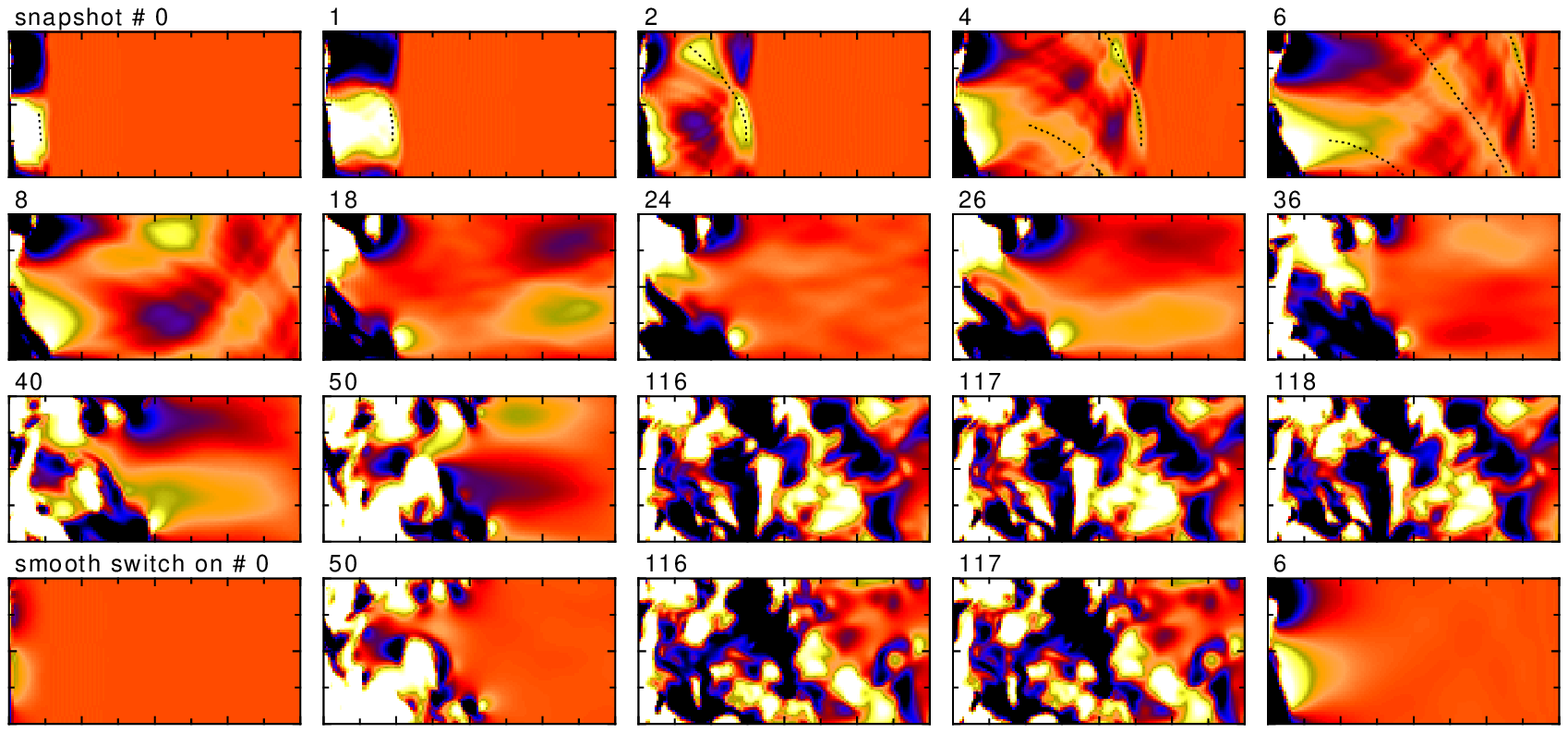}{driver_switch-on}{
Temporal evolution of a 3D MHD setup with a turbulent mixing driver at the left boundary.
The panels show different snapshots, where the top three rows belong to a run with the driving switched on instantly, and in the bottom row we show the exact same setup with the driving switched on smoothly within the time until the first snapshot.
The color indicates the vertical velocity: red represents the zero level, yellow to white indicates positive values and blue to black negative values.
The black dotted lines indicate one shock-wave front propagating through the simulation domain.
The domain is periodic in the vertical direction, so that the wave front that leaves the simulation domain through the top boundary will enter again through the lower boundary.}

The upper three rows in \fig{driver_switch-on} show the time evolution for the instant switching on of the mixing velocities at the left boundary.
This means that we apply a very strong impulse within the first time step, where we jump instantly from zero to the target velocities in snapshot \#\,0.
In the subsequent snapshots \#\,1--6 we see how the generated shock front propagates to the right boundary and actually reaches it in snapshot \#\,6.
The horizontal propagation speed stays constant.

Furthermore, we also see in the top row of \fig{driver_switch-on} that the ignited waves have not only vertical shock fronts (dotted lines in snapshots \#\,0--1), but are in fact ignited by point sources and therefore we see increasing inclination angles when this front propagates to the right (dotted lines in snapshots \#\,2--6).
Due to the periodic upper and lower boundaries, the horizontal component of the shock front will never leave the simulation domain, so that it remains visible and interacts with its symmetric (negative amplitude) counterpart, as we see in the snapshots \#\,8--36 (second row).
For example, we see how the waves exchange their places in the snapshots \#\,8 and 18, as well as \#\,26 and 36, while in between (e.g. \#\,24) one could think that these waves have finally dissipated, but that particular snapshots seems to show only the situation when both, positive and negative shock wave actually interfere and cancel each other for a short moment in time.

Even at a much later time, when the physical turbulent flow has already propagated into the full simulation domain (see snapshots \#\,116--118 in the third row of \fig{driver_switch-on}), the oscillations are still identifiable by an alternating increase and decrease of the area of the white and black patches.
The proof that these oscillations are due to the described shock waves generation can easily be given by looking at the last row of \fig{driver_switch-on}, where we show the same snapshots from the same model setup, but with a smooth switching on of the mixing driver, which shows absolutely no shock waves or spurious oscillations.
Here, the driver is switched on during the time between the initial condition and the first snapshot \#\,0.
Also a velocity damping is applied during the switch-on phase, which also needs to fade out smoothly.

Interestingly, we also see here, that even the physical part of the domain seems to be still influenced by the described switch-on effect, even though the result looks very similar to the "true" solution, c.f. the upper right part of the snapshots \#\,116--117 in the third and the last row of \fig{driver_switch-on}, where the turbulent perturbations are significantly less developed in the model that was switched on smoothly.
This should underline the importance of equilibrated initial conditions and smooth switching on of any HD, MHD, or similar simulations with a time evolution.

\section{Conclusions\label{S:conclusions}}

We present a new combined 1D stratified atmospheric profile that can serve as initial condition for HD and MHD simulations from the interior of the Sun to the upper corona.
The profile is equilibrated by dissipating initial disturbances in a 1D HD model.
With this equilibrated density, temperature, and vertical velocity profiles together, we are able to start a 3D MHD setup without any switch-on effects, such as relaxation motions, shock waves, or oscillations.
Thereby, we can directly start computing the purely physical part of the simulation, saving a lot of computing time in the production model setup.

We propose to use the obtained stratified atmosphere as reference model for inversions of spectral observations.
On the first look, these profiles might look like the chromosphere would be too thick and the transition region would be not steep enough.
But, by fact, the solar atmosphere is highly dynamic and reference profiles should not only reflect a quiet Sun situation with purely vertical field.
Instead, one should prefer such an average profile and take into account a mixture of active regions, plage and network areas, as well as quiet Sun, where the magnetic fields are usually not vertical and the average profile is then looking much more like the one we present here, as we can see from our active region model that matched well various coronal observations.

\section*{Acknowledgements} 
We used the Pencil Code\footnote{http://Pencil-Code.nordita.org/} for obtaining the results shown in this work.

\bibliographystyle{ceab}
\bibliography{Literatur}

\end{document}